\documentstyle[prb,aps,epsf,psfig,epsfig,multicol]{revtex}

\begin{document}
\sloppy

\title{\vskip0.5cm
Lattice-model study of the thermodynamic interplay of polymer crystallization and liquid-liquid demixing}
  \author{Wenbing Hu$^1$, Daan Frenkel$^1$, Vincent B. F. Mathot$^2$\\$^1$FOM
Institute for Atomic and Molecular Physics,\\ Kruislaan 407, 1098 SJ Amsterdam,
The Netherlands \\$^2$DSM Research, P. O. Box 18 Geleen, The Netherlands}

\maketitle
%\newpage
\begin{abstract}
We report Monte Carlo simulations of a lattice-polymer model that
can account for both polymer crystallization and liquid-liquid
demixing in solutions of semiflexible homopolymers. In our model,
neighboring polymer segments can have isotropic interactions that
affect demixing, and anisotropic interactions that are responsible
for freezing. However, our simulations show that the isotropic
interactions also have a noticeable effect on the freezing curve,
as do the anisotropic interactions on demixing.  As the relative
strength of the isotropic interactions is reduced, the
liquid-liquid demixing transition disappears below the freezing
curve. A simple, extended Flory-Huggins theory accounts quite well
for the phase behavior observed in the simulations.
\end{abstract}
%\newpage
\begin{multicols}{2}
\section{Introduction}
Lattice models of polymer solutions are widely used because of
their simplicity and computational convenience.
\cite{1,2,3,4,5,6,7,8}  When modeling a polymer solution, the
polymer chain occupies consecutive sites on the lattice, each site
corresponding to the size of one chain unit, while the remaining
sites correspond to solvent.

 The use of lattice models for polymer solutions dates back to
the work of Meyer \cite{1}.
Flory \cite{2} and Huggins \cite{3} showed how, using a mean-field
approximation, the lattice model yielded a powerful tool to
predict the solution properties of flexible \cite{9,10,11} and
semi-flexible \cite{7} polymers. Various refinements to the
Flory-Huggins (F-H) model have been proposed by a number of
authors (see, e.g. Refs. 4 to 6).  FH-style models
can account for liquid-liquid (L-L) phase separations with an
upper critical solution temperature (UCST) driven by the
site-to-site mixing pair interactions in polymer solutions -
however, they are ill suited to describe polymer crystallization,
i.e. liquid-solid (L-S) phase transitions.  This limitation is not
due to any intrinsic drawback of polymer lattice models as such,
but to specific choice for the polymer interactions in the
original F-H theory. In fact, the factors that lead to polymer
crystallization, i.e. interactions that favor compact packing and
stiffness of the polymer chains, can be accounted for in a lattice
model, by introducing anisotropic  interactions between adjacent
polymer bonds.~\cite{8} Clearly, in real polymer solutions, both
crystallization and phase separation can occur on cooling. While
lattice models for polymer solutions can account for both types of
phase transitions, most theoretical and simulation studies have
focused on one transition or the other, and less attention has
been paid to their interplay. Such interplay may change the
pathway of a phase transition \cite{12,13} and hence determine the
complex structure-property relationships of mixtures containing
crystallizable polymers, which has been the subject of much
experimental research dating back to Richards. \cite{14}

\begin{figure}[h]
\centering\psfig{file=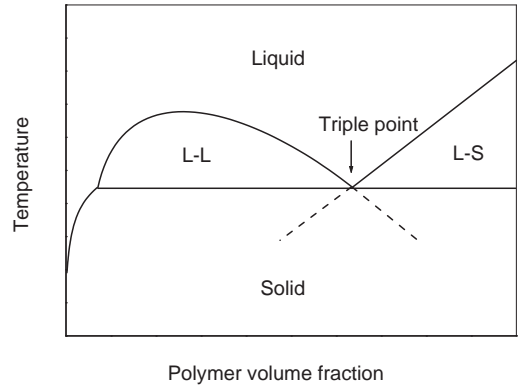, height=6 cm}
\caption{Schematic phase diagram of a binary mixture with a conventional
monotectic triple point.}
\label{fig1}
\end{figure}

  When the L-S phase-transition curve intersects the L-L coexistence curve,
both curves are terminated at the resulting triple point.
Below the triple point, the fluid phase may phase-separate into a dilute
solution and a dense crystalline phase, as depicted in
Fig.\ref{fig1}.  This combination of L-L demixing and
crystallization is often referred to as ``monotectic'' behavior and
has been observed in many experiments. \cite{10,15,16}  The
morphology of polymer crystallites appears to be sensitive to the
result of thermodynamic competition on cooling.\cite{17} Special
attention has been focused on the monotectic triple point. The
kinetic competition between L-L demixing and crystallization on
cooling in the vicinity of this triple point is an important issue
for sol-gel transition and membrane preparation. \cite{18,19,20}
On cooling through the triple point, L-L phase separation is
expected to occur before crystallization, though both phase transitions
have the same equilibrium temperatures.\cite{21} As a
consequence, the density modulation produced during the early
stage of L-L demixing may be frozen by subsequent crystallization.
\cite{22}  Such frozen-in density modulations can be a practical
way to control the metastable morphology of polymer gels and
membranes through thermally induced processes. Therefore, the
ability to predict phase diagrams of the type shown in Fig. \ref{fig1}
could be of considerable practical importance.

In this paper, we study the interplay of polymer crystallization
and L-L demixing using both mean-field theories and Monte Carlo
simulations of simple lattice models.  In particular, we pay
attention to the shift of the crystallization and L-L demixing
curves in the phase diagrams due to this interplay.

  The remainder of this paper is organized as follows: after an introductory
description of the simulation techniques, we compare the simulation results with
the relevant theoretical predictions for the L-L phase separation curve without
prior disorder-order phase transition on cooling.
Next, we discuss the simulations and mean-field calculation of 
the L-S curves and its thermodynamic
competition with L-L demixing.

\section{Simulation Techniques}
In our Monte Carlo simulations, we used a single-site-jumping
micro-relaxation model with local sliding diffusion \cite{23} to
model the time evolution of self- and mutually-avoiding polymers
in a cubic lattice with periodic boundary conditions.  In this
model, monomer displacements are allowed along both the cubic axes
and the (body and face) diagonals, so the coordination number of
each site includes all the neighboring sites along the main axes
and the diagonals, and is $q=6+8+12=26$. The single-site jumping
model with either kink generation or end-to-end sliding-diffusion
was first proposed by Larson et al.\cite{24} The kink-generation
algorithm was subsequently developed into the bond-fluctuation
model.\cite{25,26} A hybrid model combining kink generation and
sliding diffusion into one mode of chain motion, was suggested by
Lu and Yang.\cite{27} The present hybrid model considers
sliding-diffusion moves that are terminated by smoothing out the
nearest kink conformation along the chain,\cite{23} in accord with
de Gennes's picture of defect diffusion along the chain.\cite{28}
It has been verified that this model correctly reproduces both
static and dynamic scalings of short polymers in the
melt.\cite{29}

In our simulations, we consider systems containing a number of
$32$-unit polymer chains. The polymers reside in a cubic box with
$32^3$ lattice sites. The polymer concentration was varied by
changing the number of polymers in the simulation box. Monte Carlo
sampling was performed using the Metropolis method. Three
energetic parameters were used to model the intra- and
inter-molecular interactions of the polymers.  The first parameter
$E_c$ measures the energy penalty associated with having two
non-collinear consecutive bonds (a ``kink'') along the chain; it is
a measure of the rigidity of chains.  The second parameter $E_p$
measures the energy difference between a pair of parallel and
non-parallel polymer bonds in adjacent, non-bonded  positions. A
positive value of $E_p$ favors the compact packing of parallel
chain molecules in a crystal. Finally, the parameter $B$ describes
the energy penalty for creating a monomer-solvent contact. The
total change in potential energy associated with a Monte Carlo
trial move is
\begin{eqnarray}
\frac{\Delta E}{k_BT}&=& \frac{E_c \Delta c+E_p \Delta p + B \Delta m}{k_BT}
 \nonumber \\& =&( \Delta c+\Delta p \frac{E_p}{E_c} + \Delta m \frac{B}{E_c})
\frac{E_c}{k_BT},
\label{eq1}
\end{eqnarray}
where $\Delta c$ denotes the net change in the number of kinks,
$\Delta p$ is the change in the number of non-parallel adjacent
bonds, and $\Delta m$ measures the change in the number of
monomer-solvent contacts. $k_B$ is the Boltzmann constant and $T$
is the temperature.  As shown in Eq. \ref{eq1}, three
dimensionless parameters control the acceptance probability of
Monte Carlo trial moves: $B/E_c$ is the term that dominates the
L-L demixing temperature but no effect at all on the freezing of
the pure polymer system. In contrast, $E_p/E_c$ completely
determines the freezing temperature of the pure polymer system,
but it has only a slight effect on the demixing temperature. In
fact, from  Eq. \ref{eq5} below, it follows that, the critical
demixing temperature is approximately a factor of $q$  higher in the
case $E_p=0$ and $B/E_c\ne 0$ than in the case where the values of
$B$ and $E_p$ are interchanged. In what follows, $E_c/(k_BT)$ is
used as a measure of the (inverse) temperature of the system. If
$E_c$ is much larger than $B$ and $E_p$ , the polymer chains
behave as almost rigid rods. In contrast, if $E_c=0$, the polymers
are fully flexible. In what follows, we chose $E_p/E_c=1$ as a
value typical for semiflexible chains. The choice of the value of
$B/E_c$ (and thereby the L-L demixing region) is discussed in the
following sections. In our simulations, we lowered the temperature
by increasing the value of $E_c/(k_BT)$ from zero in steps of
$0.002$. At each step, the total number of trial moves was $500$
MC cycles, where one Monte Carlo cycle (MC cycle) is defined as
one trial move per monomer. The first $400$ MC cycles at each
temperature were discarded for equilibration, after which samples
were taken once per MC cycle and averaged. This process
corresponds to a slow cooling of the sample system.

The most direct way to establish the equilibrium phase diagram of
this model system would be to compute the free energy of all
phases. Here, we follow a different route: we attempt to locate
the equilibrium phase-transition temperatures during the dynamic
cooling process. However, rapid cooling may lead to a significant
supercooling mainly due to the presence of a free-energy barrier
for homogeneous nucleation. This is particularly true in dilute
solutions and small systems. In order to identify the correct
equilibrium coexistence curves in a dynamic cooling scheme,
supercooling should be eliminated as much as possible. To this
end, we introduced one solid layer of terraced substrate formed by
extended chains, as shown in Fig. \ref{fig2}A.  These terraces can
induce heterogeneous nucleation with a very small free-energy
barrier. On such a large, terraced substrate, layer-by-layer
crystal growth can take place directly, thereby obviating the need
for homogeneous nucleation. In order to increase the accuracy of
the method near the onset of the phase transition, we monitored
the properties of the system during successive blocks of $500$ MC
cycles. If, during  such a block, we found evidence for the onset
of a phase transition, we kept the temperature constant for a
number of subsequent blocks, until no further drift in the system
properties was observed.

     On cooling, the degree of order in the sample system can be traced by the
Flory ``disorder'' parameter, defined as the mean fraction of
non-collinear connections of two consecutive bonds along the
chains.  On the cubic lattice, where $24$ out of $25$ directions
for the connection to the next bond are non- collinear, the
high-temperature limit of the disorder parameter is $0.96$. The
degree of demixing of the system can be monitored by tracing the
value of a ``mixing'' parameter, defined as the mean fraction of the
sites around a chain unit, that are occupied by solvent. Our
estimates of the onsets of phase transitions are based on the
averaged results of  five independent cooling processes
characterized by the same energy parameters, but different seeds
for the random-number generation.

\end{multicols}
\begin{multicols} {2}
\begin{figure}[t]
\centering\psfig{file=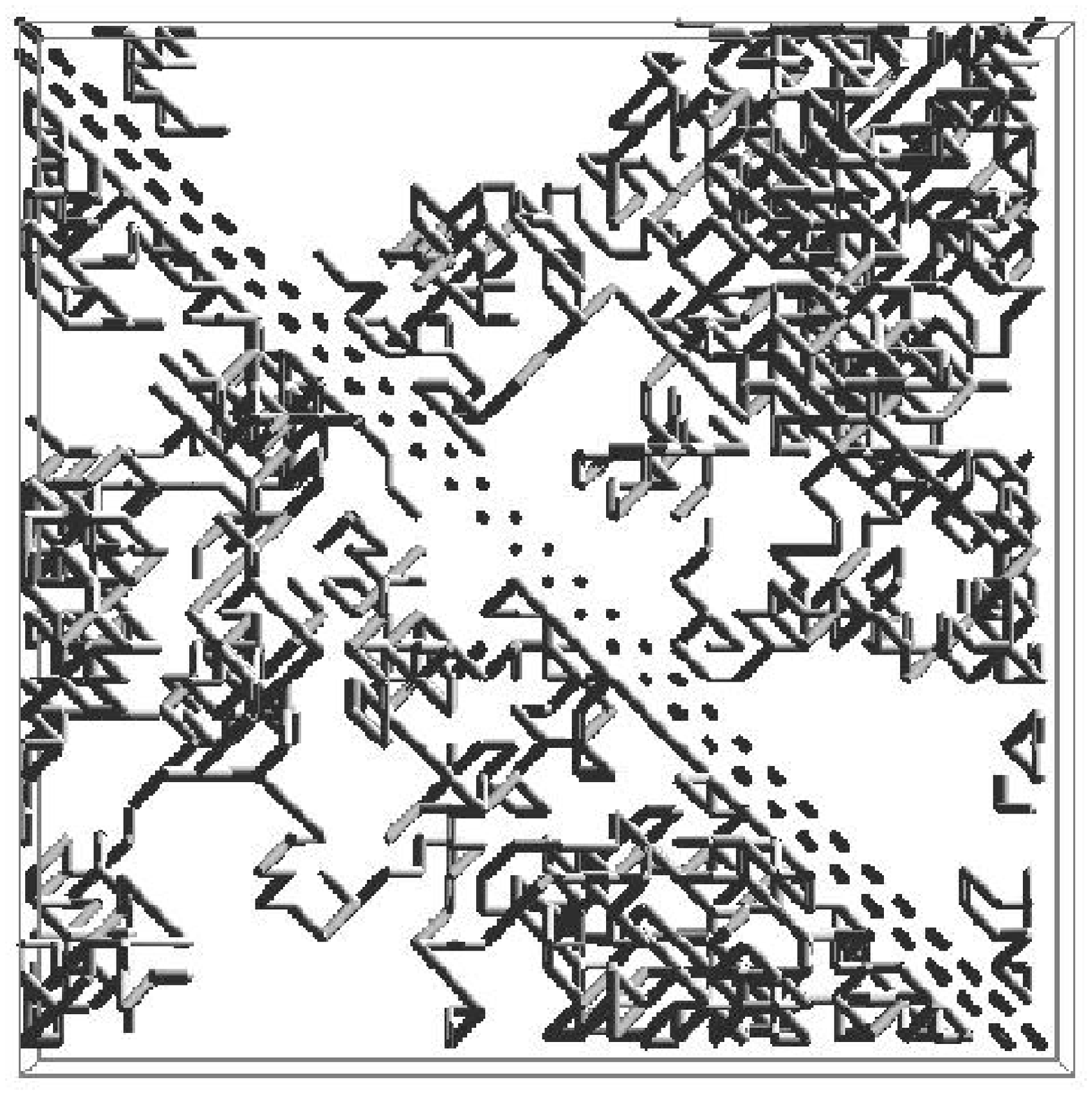, height=6 cm} \\ \textbf{A.} \\
\epsfig{file=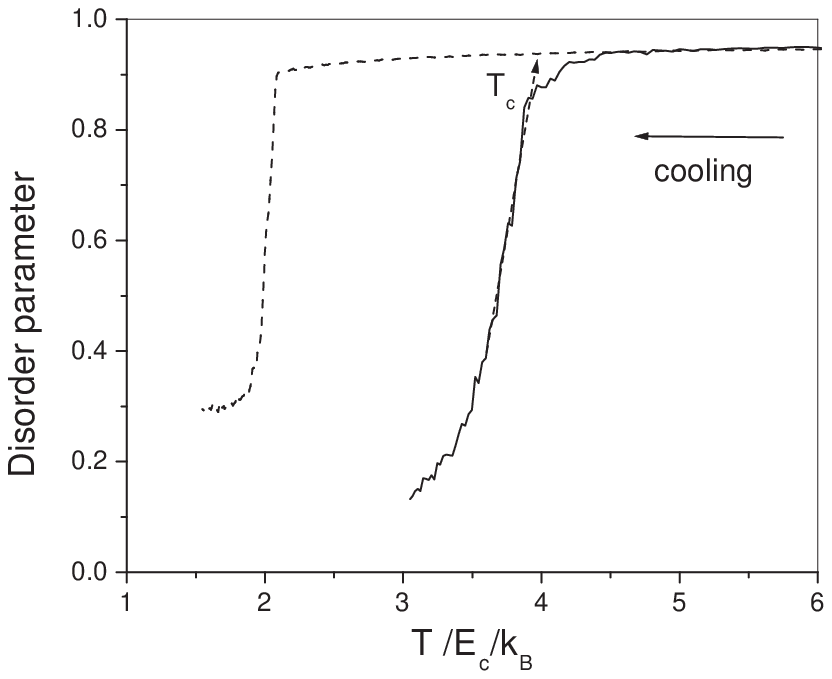, height=6 cm} \\ \textbf{B.}  
\end{figure}
\begin{figure}[t]
\centering\epsfig{file=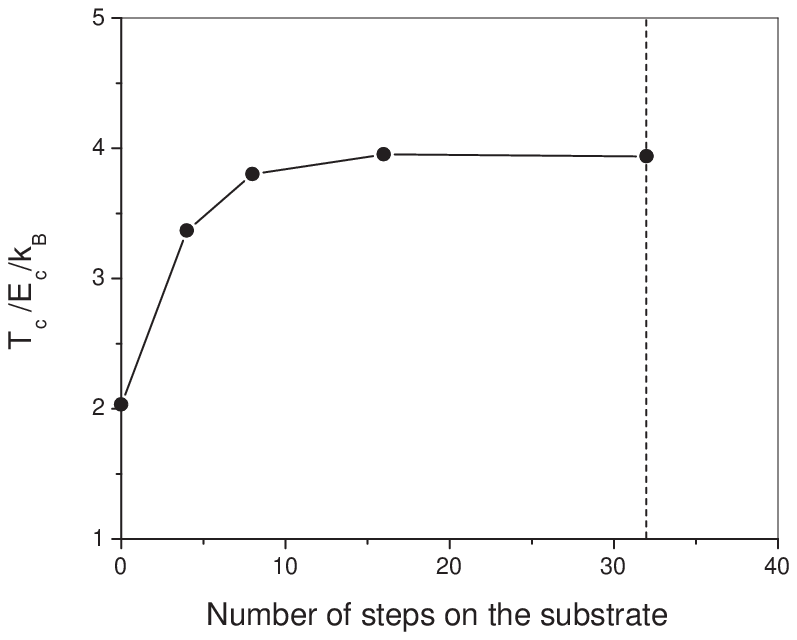,height=6 cm} \\ \textbf{C.} \\ 
\epsfig{file=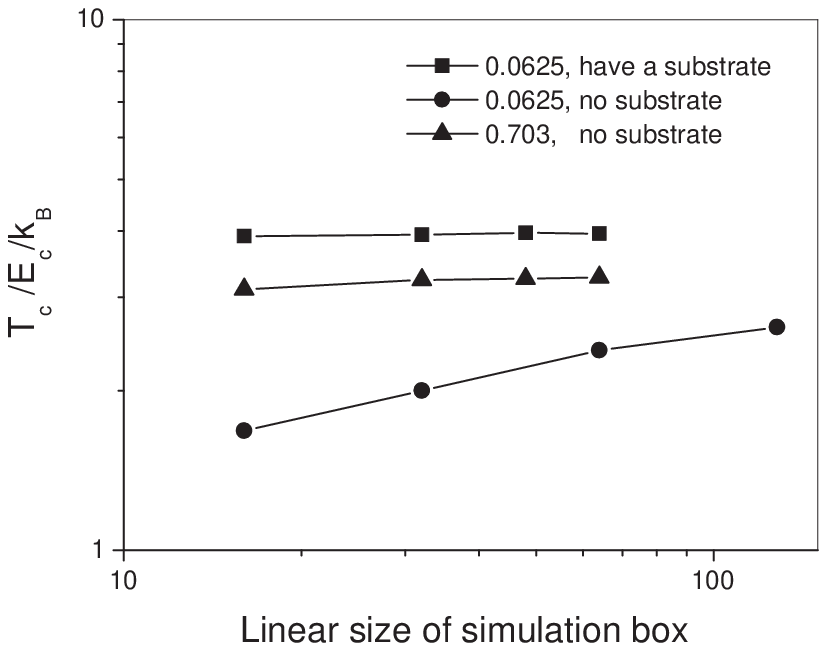,height=6 cm} \\  \textbf{D.} 
\end{figure}

\end{multicols}

\begin{figure}[h]
\caption{Effect of a terraced substrate on the onset of
crystallization upon cooling. The figures shown above were
obtained for a solution of model polymers with length $r=32$, at a
volume fraction $\phi=0.0625$, with  $E_p/E_c=1$ and $B/E_c=0$.
(A) Snapshot of an athermal sample system containing one layer of
terraced substrate formed by extended chains, that are not
included in the polymer volume fraction. Viewing along the
extended chains. (B) Disorder-parameter cooling curves for the
sample systems with a terraced substrate on cooling (solid line)
and under the absence of a seed on cooling (dashed line). The
arrow indicates the onset of phase transition. (C) Step-size
dependence of the onset of crystallization on cooling. (D)
Finite-size scaling of the onset of crystallization on cooling for
the sample systems with denoted concentrations. All error bars are
smaller than the symbols. The segments are drawn as a guide to the
eye.}
\label{fig2}
\end{figure}

\begin{multicols}{2}

As can be seen from Fig. \ref{fig2}B, the presence of a terraced
substrate significantly decreases the kinetic delay on cooling for
polymer crystallization from a dilute solution. The onset of
crystallization induced by the terraced substrate becomes
insensitive to the number of steps on the substrate when this
number is larger than $8$, see Fig. \ref{fig2}C. One might expect
that more steps on the substrate would cause the substrate to
adsorb more chains. The fact that the phase-transition temperature
becomes insensitive to the number of steps (here, and in what
follows, we use $32$ steps), suggests that pretransitional
adsorption has a negligible effect on the apparent
phase-transition temperature. In contrast, if no ``template'' is
present, the onset of crystallization from a dilute solution,
depends strongly on the system size. This effect is probably due
to the volume dependence of the homogeneous nucleation rate. It
can be completely eliminated by the introduction of a terraced
substrate,  as demonstrated in Fig. \ref{fig2}D.

   In the following sections, we first consider the case that $E_p/E_c$
is zero and hence no crystallization can take place, while $B/E_c$
is large enough to induce L-L demixing on cooling. Next, we switch on 
$E_p/E_c$. This allows us to study  a phase
diagram that exhibits both L-L demixing and freezing.

\section{Results and Discussion}

\subsection{Liquid-liquid demixing without crystallization}
If both $B/E_c$ and $E_p/E_c$ are zero, the model only takes
excluded-volume interactions and the temperature dependence of
chain flexibility into account. Even in this case, the polymer
solution may exhibit a disorder-order phase transition on cooling.
\cite{8,30} This transition is not, strictly speaking, a freezing
transition but rather an isotropic-nematic phase transition: it is
induced by the increase in anisotropic excluded volume
interactions between polymer chains, due to the increase in chain
rigidity on cooling. \cite{7,31} This transition has recently been
studied extensively by Weber et al. \cite{32}

If we increase the value of $B/E_c$ while keeping $E_p/E_c$ equal
to zero, we should reach a point above which L-L demixing occurs
prior to the isotropic- nematic phase transition on cooling.

We focused our attention on the L-L demixing with values of
$B/E_c$ beyond that critical value, and kept track of the
``mixing'' parameter on cooling.  As the dense liquid phase wets
the terraced substrate, the onset temperature of L-L demixing
induced by such a substrate should be a good approximation to the
equilibrium phase separation temperature.  A tentative binodal
curve can thus be obtained in simulations to compare with the
predictions of mean-field theories.

\begin{figure}[h]
\centering\epsfig{file=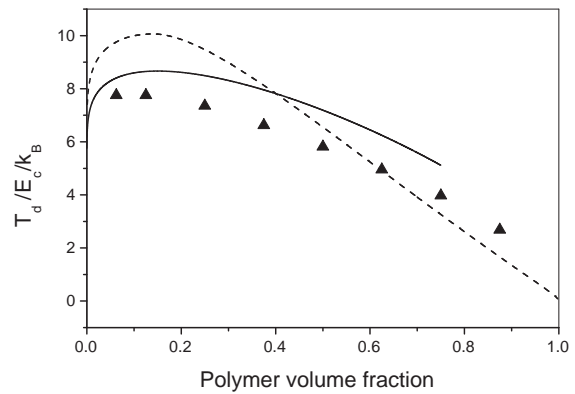, height=6 cm}
\caption{Liquid-liquid coexistence curves ($T_d$) of the sample system with
$E_p/E_c=0$ and $B/E_c=0.25$.  The solid line is calculated from the classical
Flory-Huggins free-energy expression for polymer solutions, and the dashed line
is calculated from the second-order expansion of the mixing free energy in
lattice-cluster theory. The triangles are the onsets of liquid-liquid demixing
induced by a terraced substrate on cooling. The error bars are smaller than the
symbols.}
\label{fig3}
\end{figure}

Figure \ref{fig3} shows the binodal curves for the sample systems
with $E_p/E_c=0$ and $B/E_c=0.25$.  The binodal curve can be
estimated from the condition of equal chemical potential of the
coexisting phases, using the Eqn.~\ref{eq2}, the F-H expression
for the mixing free-energy.
\begin{eqnarray}
\frac{\Delta f_{mix}}{k_BT}=(1- \phi )\text{ln}(1- \phi
)+\frac{\phi}{r}\text{ln}(\phi ) + \phi (1- \phi )
\frac{(q-2)B}{k_BT}, \label{eq2}
\end{eqnarray}
where $\phi$ is polymer volume fraction, $r$ is the chain length,
and $q$ the lattice-coordination number. As can be seen from
Fig.~\ref{fig3}, the theoretical predictions show a small but
constant deviation from the simulation results.

Yan et al.~\cite{33}  have shown that a second-order
lattice-cluster theory may provide a better description of the
critical point of the binodal curve obtained in computer
simulations. To second order, the mixing free-energy change per
lattice site is \cite{6}
\begin{eqnarray}
\frac{\Delta f_{mix}}{k_BT}&=&(1- \phi )\text{ln}(1- \phi
)+\frac{\phi}{r}\text{ln}( \phi ) \nonumber \\ &&-\frac{1}{2}q
\epsilon \phi ^2+C_0+C_1 \epsilon +C_2 \epsilon^2, \label{eq3}
\end{eqnarray}
where $\epsilon =2B/(k_BT)$. Explicit expressions for $C_0$, $C_1$
and $C_2$  in terms of $\phi$, $q$ and $r$ are given in Ref. 6.
When we compare the predictions of the second-order
lattice-cluster theory with our simulations, (dashed curve in Fig.
\ref{fig3}) we find that this theory does not lead to better
agreement with the simulation data, except perhaps at high polymer
concentrations. It should be noted that, for very long polymer
chains, the lattice cluster theory may predict more than one
critical point. \cite{34} Hence, the predictions of this theory
should be viewed with some caution. \cite{35}

\subsection{Polymer crystallization and its interplay with liquid-liquid demixing}

When we set $B/E_c=0$ and $E_p/E_c=1$, L-L demixing is preempted
by freezing. In fact, an estimate based on mean-field theory
(Eqn.~\ref{eq4a} below) indicates that, for these parameter
values, the freezing temperature of the pure polymer is a factor
four higher than the critical demixing temperature. We assume that
the onset of crystallization induced by the terraced substrate,
yields a good approximation for the equilibrium melting
temperature. It is this temperature that we subsequently compare
with the corresponding prediction of mean-field theory.

The mean-field expression for the partition function of the
disordered polymer solution is given by~\cite{8}:
\begin{equation}
Z=(\frac{n}{n_1})^{n_1}(\frac{n}{n_2})^{n_2}(\frac{q}{2})^{n_2}z_c^{(r-
2)n_2}\text{e}^{(1-r)n_2} z_p^{(r-1)n_2} z_l^{rn_2}, \label{eq4}
\end{equation}
where \\
$z_c=1+(q-2)\text{exp}(-\frac{E_c}{k_BT})$,\\
$z_p=\text{exp}[-\frac{q-2}{2}(1-\frac{2(r-1)n_2}{qn}) \frac{E_p}{k_BT}]$,\\
$z_l=\text{exp}(-\frac{n_1}{n} \frac{(q-2)B}{k_BT})$\\
$n_1$ denotes the number of sites occupied by the solvent, $n_2$
the number of chains, each having $r$ units, and $n=n_1+r n_2$. We
note that, in the above expression, we have corrected an error in
the expression for the partition function given in Ref. 8. The
corresponding expression for the free-energy density (i.e. the
Helmholtz free energy per lattice site) is
\begin{eqnarray}
\frac{f(\phi)}{k_BT}&=&(1-\phi)\ln(1-\phi)+\frac{\phi}{r}\ln\phi
\nonumber\\
&& +\phi(-\frac{\ln(qr/2)}{r} -(1-2/r)\ln z_c + (1-1/r) 
\nonumber\\
&&  +(q-2)\frac{B}{k_BT}+(1-1/r)\frac{q-2}{2}\frac{E_p}{k_BT})\nonumber\\
&& -\phi^2\left((q-2)\frac{B}{k_BT}+(1-1/r)^2\frac{q-2}{q}\frac{E_p}{k_BT}\right)\label{eq4a}
\end{eqnarray}

We assume that the pure polymer crystal is in its fully ordered
ground state and that the partition function of this state is
equal to one. In a pure polymer system, melting takes place at the
point where the free energies of the crystal and the melt cross.
For polymer solutions, the freezing curve can be computed by
imposing that the chemical potential of the polymers in crystal
and solution are equal, i. e. $\mu^c-\mu^0=\mu^s-\mu^0$, where
$\mu^0$ is the chemical potential of polymers in the ground state.
As the free energy of the crystal phase is assumed to be equal to
zero, the chemical potential of the polymers in that phase is also
equal to zero.The chemical potential of the polymers in solution
is $\mu^s=\partial F^s/\partial n_2$. Thus by solving the equation
$\partial \text{ln}Z^s/\partial n_2=0$ by iteration, we can obtain
the equilibrium melting temperature.

    Starting the calculation from Eq. \ref{eq4a}, the F-H expression for the
mixing free-energy change becomes
\begin{eqnarray}
\frac{\Delta f_{mix}}{k_BT}&=&(1- \phi )\text{ln}(1- \phi
)+\frac{\phi}{r}\text{ln}(\phi ) \nonumber \\ &&+ \phi(1- \phi )(q-2)
\left(\frac{B}{k_BT}+\frac{1}{q}(1-\frac{1}{r})^2\frac{E_p}{k_BT}\right).
\label{eq5}
\end{eqnarray}
The binodal L-L curves can be separately estimated without
the consideration of L-S curves.

\begin{figure}[h]
\centering{\epsfig{file=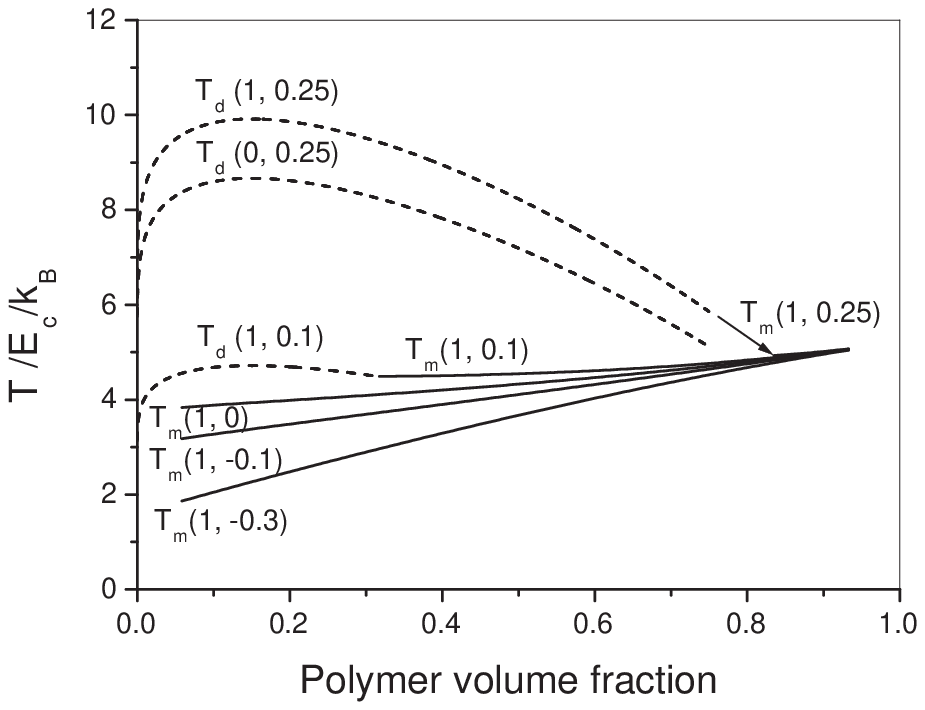,height=6 cm} \\ \textbf{A.}
\\          \epsfig{file=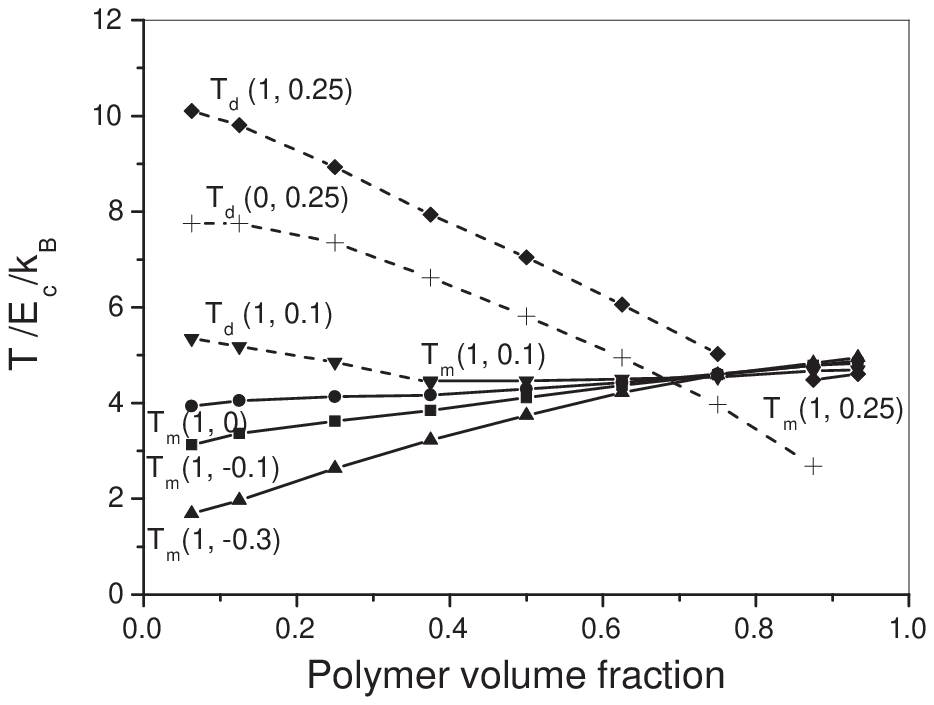,height=6 cm} \\ \textbf{B.}}
\caption{Liquid-liquid demixing curves (denoted as $T_d$) and liquid-solid
transition curves (denoted as $T_m$) for the sample system with
variable energy parameter settings (denoted as $T(E_p/E_c,
B/E_c)$).  (A) Theoretical curves calculated from Eq. \ref{eq4}
with $q_{eff}=q-2$ (Flory-Huggins approach). Note that changing
$E_p/E_c$ from 1 to 0, leads to a 10\% decrease in $T_d$. In
contrast, lowering $B/E_c$ by 0.15 reduced $T_d$ by more than
50\%. An arrow indicates the position of possible triple point;
(B) Onsets of phase transitions induced by a terraced substrate on
cooling. The error bars are smaller than the symbols, and the
segments are drawn to guide the eye.}
\label{fig4}
\end{figure}

In Figs. \ref{fig4}A and \ref{fig4}B, we compare the mean-field predictions
for the phase diagram with the simulation data. In view of the
simplicity of the mean-field theory, the agreement between theory
(without adjustable parameters) and the simulation data, is
gratifying.

According to Eq. \ref{eq5}, we should expect that a positive value
of $E_p/E_c$ will increase the L-L demixing temperature.  This is
precisely the behavior observed in Fig. \ref{fig4}, where the L-L
demixing curve of the sample system with $B/E_c=0.25$ shifts up
when the value of $E_p/E_c$ changes from zero to one. By carefully
choosing the parameters, such as $B/E_c=0.1$, we can ``tune'' the
relative strength of the tendencies to crystallize and to demix,
and observe the intersection of the L-S and L-L curves.

Although a change in the value of $B/E_c$ cannot change the
freezing temperature of pure polymers, it can change the L-S
coexistence curve of polymer solutions. The reason is that a
poor solvent favors phase separation (be it L-L or L-S).

 However, in the simulations, we observed that the L-S curves cross not only at
$\phi=1$ but also at a second point near $\phi=0.73$. This
crossing point is not related to the presence of the terraced
substrate, as it has also been observed in the absence of such a
template.\cite{8} Possibly, this failure of the simple mean-field
theory is due to the rather naive way in which it accounts for the
effective coordination of monomers. We point out that, in our
estimate, we have assumed that the effective coordination number
is equal to $q-2$. But, in more sophisticated theoretical
descriptions, $q_{eff}$ (as in Eq.~\ref{eq3}) is, itself,
concentration dependent.

\begin{figure}[h]
\centering\psfig{file=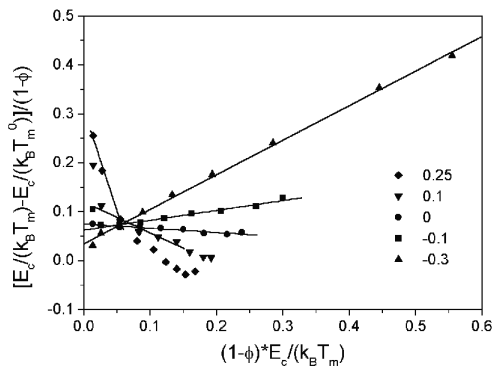, height=6 cm}
\caption{Rescaled data in Fig. \ref{fig4}B for the onsets of crystallization induced
by a terraced substrate on cooling, according to the formula of
Eq. \ref{eq6} with an approximation of $E_c/(k_BT_m^0)=0.2$.  The
solid lines are the results of linear regression of those data
points. The meaning of the symbols is the same as those in figure
~\ref{fig4}B.}
\label{fig5}
\end{figure}

  Flory has proposed a semi-empirical relationship between the melting point and
the concentration of polymers in solutions, \cite{30} as given by
\begin{equation}
\frac{1}{T_m} -\frac{1}{T_m^0} = \frac{k_B}{\Delta h_u}[1-\phi-
\frac{q_{eff}B}{k_BT_m} (1- \phi )^2],
\label{eq6}
\end{equation}
where $T_m^0$ is the equilibrium melting point of bulk polymers,
$\Delta h_u$ is the heat of fusion per chain-unit. The predictions
of Eq. \ref{eq6} for the melting-point depression upon dilution
have been verified by several experimental measurements at both
high and low concentration ends. \cite{31,32} The linear
relationship predicted by Eq. \ref{eq6} does hold for those
simulations where L-L demixing does not occur (see Fig.
\ref{fig5}). According to Eq. \ref{eq6}, the values of
$-q_{eff}B/\Delta h_u$ and $E_c/\Delta h_u$ can be obtained
respectively from the slope and the intercept of the freezing
``line''. We found that $-q_{eff}B/\Delta h_u$ depends nearly
linearly on $B/E_c \times E_c/\Delta h_u$. In addition, $\Delta
h_u/E_c$ varies linearly with $B/E_c$. Assuming that both
relations are, in fact, linear, we find: $q_{eff}=54.0$ and
$\Delta h_u=41.0B+13.0E_c$ respectively. The latter result implies
a microscopic coupling between L-L demixing and polymer
crystallization, consistent but not identical with the previous
study. \cite{8}

In this paper, we have addressed the equilibrium freezing and
demixing curves of lattice polymers. In subsequent work, we shall
address the effect of the interplay between  demixing and freezing
on the kinetics of the phase transformation.

\textbf{Acknowledgement} W.H. acknowledges helpful discussion with
Dr. Mark Miller. This work was financially supported by DSM
Company. The work of FOM institute is part of the research program
of the ``Stichting voor Fundamenteel Onderzoek der Materie (FOM)'',
which is financially supported by the ``Nederlandse organisatie
voor Wetenschappelijk Onderzoek (NWO)''.

%\newpage

\end{multicols}


\begin{thebibliography}{38}
\bibitem{1} K.H. Meyer, Z. Phys. Chem. (Leipzig) B\textbf{44}, 383(1939).
\bibitem{2} P.J. Flory, J. Chem. Phys. \textbf{10}, 51(1942).
\bibitem{3} M.L. Huggins, Ann. N. Y. Acad. Sci. \textbf{43}, 1(1942).
\bibitem{4} R. Koningsveld, L.A. Kleintjens, Macromolecules \textbf{4}, 637(1971).
\bibitem{5} M.G. Bawendi, K.F. Freed, J. Chem. Phys. \textbf{88}, 2741(1988).
\bibitem{6} D. Buta, K.F. Freed, and I. Szleifer J. Chem. Phys. \textbf{112},
6040(2000).
\bibitem{7} P.J. Flory, Proc. R. Soc. London, Ser. A \textbf{234}, 60(1956).
\bibitem{8} W.-B. Hu, J. Chem. Phys. \textbf{113}, 3901(2000).
\bibitem{9} E.A. Guggenheim, \textit{Mixtures} (Clarendon, Oxford, 1952).
\bibitem{10} P.J. Flory, \textit{Principles of Polymer Chemistry} (Cornell
University Press, Ithaca, NY, 1953).
\bibitem{11} I. Prigogine, \textit{The Molecular Theory of Solution} (Amsterdam,
1957).
\bibitem{12} P. R. ten Wolde, D. Frenkel, Science \textbf{277}, 1975(1997).
\bibitem{13} V. Talanquer, D. W. Oxtoby, J. Chem. Phys. \textbf{109}, 223(1998).
\bibitem{14} R.B. Richards, Trans. Faraday Soc. \textbf{42}, 10(1946).
\bibitem{15} L. Aerts, H. Berghmans, and R. Koningsveld, Makromol. Chem.
\textbf{194}, 2697(1993).
\bibitem{16} X.W. He, J. Herz, and J.M. Guenet, Macromolecules \textbf{20},
2003(1987).
\bibitem{17} P. Schaaf, B. Lotz, J. C. Wittmann, Polymer \textbf{28}, 193(1987).
\bibitem{18} H.K. Lee, A.S. Myerson, and K. Levon, Macromolecules \textbf{25},
4002(1992) and ref. therein.
\bibitem{19} J.M. Guenet, Thermochimica Acta \textbf{284}, 67(1996).
\bibitem{20} H. Berghmans, R. De Cooman, J. De Rudder, and R. Koningsveld
Polymer \textbf{39}, 4621(1998).
\bibitem{21} R. Koningsveld, Ph.D. Thesis, University of Leiden, 1967.
\bibitem{22} N. Inaba, K. Sato, S. Suzuki, T. Hashimoto, Macromolecules
\textbf{19},1690(1986).
\bibitem{23} W.-B. Hu, J. Chem. Phys. \textbf{109}, 3686(1998).
\bibitem{24} R. G. Larson, L. E. Scriven, H. T. Davis, J. Chem. Phys.
\textbf{83}, 2411(1985).
\bibitem{25} I. Camesin, K. Kremer, Macromolecules \textbf{21}, 2819(1988).
\bibitem{26} H. P. Deutsch, K. Binder, J. Chem. Phys. \textbf{94}, 2294(1991).
\bibitem{27} J.-M. Lu, Y.-L. Yang, Sci. China Ser. A \textbf{36}, 357(1993).
\bibitem{28} P.-G. de Gennes, J. Chem. Phys. \textbf{55}, 571(1971).
\bibitem{29} W.-B. Hu, J. Chem. Phys. \textbf{115}, 4395(2001).
\bibitem{30} A. Baumgaertner, J. Chem. Phys. \textbf{84}, 1905(1986).
\bibitem{31} M. Dijkstra, D. Frenkel, Phys. Rev. E \textbf{51}, 5891(1995).
\bibitem{32} H. Weber, W. Paul and K. Binder, Phys. Rev. E \textbf{59},
2168(1999).
\bibitem{33} Q. Yan, H. Liu, and Y. Hu, Macromolecules \textbf{29}, 4066(1996).
\bibitem{34} B. Quinn, P.D. Gujrati, J. Chem. Phys. \textbf{110}, 1299(1999).
\bibitem{35} K.F. Freed, J. Dudowicz, J. Chem. Phys. \textbf{110}, 1307(1999).
\bibitem{36} P. J. Flory, J. Chem. Phys. \textbf{17}, 223(1949).
\bibitem{37} L. Mandelkern, \textit{Crystallization of Polymers} (McGraw-Hill
Book Co., NY, 1964), p. 38.
\bibitem{38} A. Prasad and L. Mandelkern, Macromolecules \textbf{22},914(1989).

\end{thebibliography}
\end{document}